\documentclass[sigconf]{acmart}

\usepackage[utf8]{inputenc}
\usepackage{amsmath,amsthm}
\usepackage{tabto}
\usepackage{algorithm}
\usepackage[noend]{algpseudocode}
\usepackage{xspace}
\usepackage{paralist}
\usepackage{url}
\usepackage{graphicx,subfigure}
\usepackage{tabularx,comment}
\usepackage{dsfont}

\newcommand{\Fig}[1]{Fig.~\ref{fig:#1}}

\newcommand{\Sec}[1]{Sec.~\ref{sec:#1}}

\newcommand{\Eq}[1]{(\ref{eq:#1})}

\newcommand{\EE}{\mathds{E}}
\newcommand{\wb}{\mathbf{w}}
\newcommand{\ub}{\mathbf{u}}
\newcommand{\vb}{\mathbf{v}}

\begin{document}

\copyrightyear{2023}
\acmYear{2023}
\acmConference[ICMLC '23]{International Conference on Machine Learning and Computing}{February 2023}{Zhuhai, China}



\title[Unexpectedly useful\dots]{Unexpectedly Useful: Convergence Bounds\\And Real-World Distributed Learning}

\author{Francesco Malandrino}
\affiliation{%
  \institution{CNR-IEIIT and CNIT}
\city{Torino}\country{Italy}
}
\author{Carla Fabiana Chiasserini}
\affiliation{%
  \institution{Politecnico di Torino, CNR-IEIIT, and CNIT}
\city{Torino}\country{Italy}
}

\renewcommand{\shortauthors}{F. Malandrino, C. F. Chiasserini}

\pagestyle{plain}

\begin{abstract}
Convergence bounds are one of the main tools to obtain  information  on the performance of a distributed machine learning task, before running the task itself. In this work, we perform a set of experiments to assess to which extent, and in which way, such bounds can predict and improve the performance of real-world distributed (namely, federated) learning tasks. We find that, as can be expected given the way they are obtained, bounds are quite loose and  their relative magnitude reflects the training rather than the testing loss. More unexpectedly, we find that some of the quantities appearing in the bounds  turn out to be very useful to identify the clients that are most likely to contribute to the learning process, without requiring the disclosure of {\em any} information about the quality or size of their datasets. This suggests that further research is  warranted on the ways -- often counter-intuitive -- in which convergence bounds can be exploited to improve the performance of  real-world distributed learning tasks.

\end{abstract}

\maketitle

\section{Introduction\label{sec:intro}}

It would be hard to overstate the importance of machine learning (ML) for a growing number of aspects of technology and, indeed, of our daily lives. Furthermore, owing to the growing complexity of the learning tasks to perform, to the ever-increasing amount of resources they require, and to the need to keep data local, a lot of today's learning is {\em distributed}, i.e., it requires the cooperation of multiple {\em learning nodes}, leveraging the help of a {\em learning server}.

\begin{figure}[t] \centering
\includegraphics[width=.8\columnwidth]{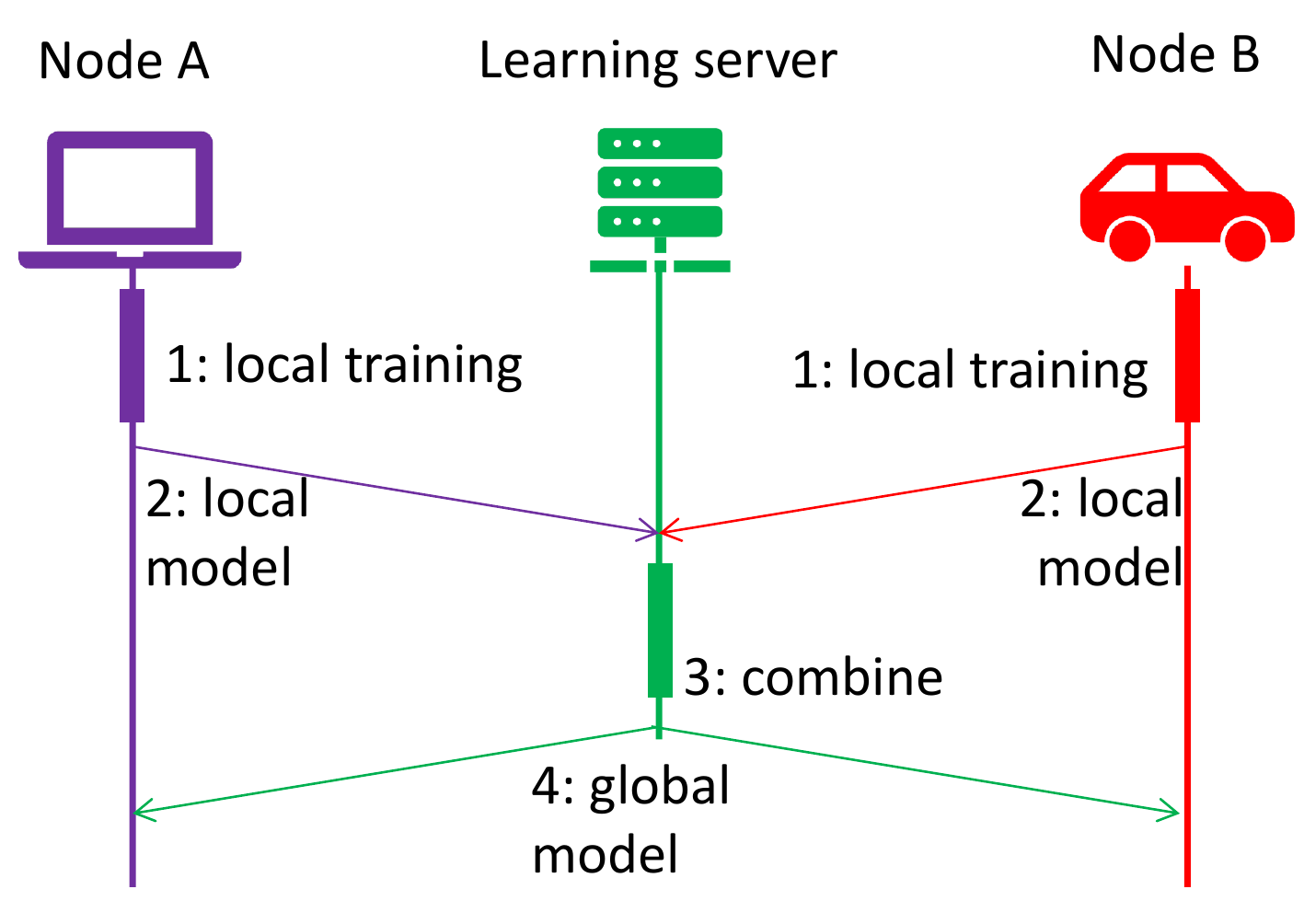} 
\caption{Main steps of each 
iteration of the federated learning paradigm:
learning nodes train their local model (1) and send the local
parameters to the server (2); the server performs a (weighted) averaging
of the model (3) and sends the global parameters back to the learning
nodes (4). 
\label{fig:sequence}} %caption 
\vspace{-4mm}
\end{figure}

A prominent example of distributed learning is represented by the Federated Learning (FL) paradigm, which operates~\cite{konen2015federatedOptimization} by performing five main steps, as summarized in \Fig{sequence}:
\begin{enumerate}
\item each learning node trains a {\em local} model, leveraging on-device data;
\item after one or more local epochs, learning nodes send their current model to the server;
\item the server creates a {\em global} model by combining, e.g., averaging~\cite{konen2015federatedOptimization}, the local models it receives;
\item the server sends the global model back to the learning nodes;
\item the learning nodes replace their local models with the global one, and resume training from step~1 above.
\end{enumerate}

In FL -- as in all types of distributed optimization -- the overall performance of learning chiefly depends upon three factors~\cite{malandrino2022network,malandrino2022energy}:
\begin{enumerate}
    \item how long each iteration (step~1 in \Fig{sequence}) takes;
    \item how much network delay (steps~2 and~4 in \Fig{sequence}) is incurred;
    \item how much the learning progresses at each iteration, hence, how many iterations are needed.
\end{enumerate}
The first two factors are widely studied, comparatively well understood, and relatively easy to estimate with a good level of accuracy. The third factor, instead, is much harder to assess; indeed, how well a learning model (e.g., a deep neural network (DNN)) can learn depends upon many factors, several of which are unknown {\em a priori}.

The most promising efforts towards modeling and estimating the progress of learning tasks focus on {\em convergence bounds}, i.e., upper bounds on the {\em loss} achieved by a given model by a certain training epoch. Such bounds may account for features of the model being trained (e.g., the number of parameters therein), the loss function (e.g., its smoothness), and the datasets being learned from (e.g., their size). Since they establish upper bounds on the loss, works on convergence analysis must account for the {\em worst-case} scenario, i.e., they describe the behavior of the model under the most unfavorable possible conditions.

In this paper, we aim at bridging the gap between theoretical works on convergence and real-world distributed learning. 
Our main contributions are twofold:
\begin{itemize}
    \item first, we assess how accurately convergence bounds capture the qualitative {\em and} quantitative behavior of concrete distributed ML;
    \item second, we find that, while the bounds themselves have a loose relationship with the actual loss evolution, the quantities needed to compute the bounds can identify the learning nodes where local iterations yield the most substantial learning  improvement~\cite{client-selection,imteaj2020fedar,zaw2021energy,malandrino2021federated}.
\end{itemize}
The latter aspect is linked to the problem of {\em selecting} the nodes  that can best contribute to the distributed training~\cite{malandrino2021federated}.
%, whilst reducing the overhead~\cite{malandrino2021federated}.

In the remainder of this paper, 
\Sec{ref} describes the convergence bounds we consider as a reference and our experimental setup, while \Sec{results} presents our experimental analysis and discusses our main findings. Finally, \Sec{conclusion} concludes the paper and sketches directions for future research.

\section{Reference bounds and experimental setup}
\label{sec:ref}

%In the following, we discuss the convergence results we compare against, as well as our experimental setup.

\subsection{Convergence bounds}

Among the many works dealing with   distributed learning convergence, we take~\cite{li2019convergence} as a reference. The main reason for our choice is that the bounds presented in~\cite{li2019convergence} account for multiple aspects of the learning scenario, hence, they are (i) more suited to assess the impact of each factor, and (ii) potentially, tighter.

Under the assumptions that all learning nodes participate in the learning process, they are equally weighted, and  one local epoch is performed for each FL iteration, \cite{li2019convergence}~proves that the difference between  expected loss~$\EE\left[F(t)\right]$ at iteration~$t$ and  minimum loss~$F^*$ is given by:
\begin{equation}
\label{eq:bound}
\frac{8L/\mu}{(t-1+8L/\mu)}\left(\frac{16G^2}{\mu}+4L\EE\lVert\wb_1-\wb^*\rVert\right)\,.
\end{equation}
In \Eq{bound}, bold letters  denote vectors; also,
\begin{itemize}
\item $\mu$ is a non-negative quantity such that  loss function~$F$ is $\mu$-strongly convex;
\item $L$ is a non-negative quantity such that  loss function~$F$ is $L$-smooth;
\item $G$ is a non-negative quantity such that the squared norm of the gradients of loss function~$F$ is bounded by~$G^2$.
\end{itemize} 
Recall that, as reported in~\cite{li2019convergence}, a loss function,~$F$, is $\mu$-strongly convex if there exists a quantity~$\mu\geq 0$ such that, for any possible model parameters~$\ub$ and~$\vb$,
\begin{equation}
\label{eq:lsmooth}
F(\ub)  \leq F(\vb) + (\ub- \vb)^T \nabla F(\vb) + \frac{L}{2} \| \ub - \vb\|_2^2\,.
\end{equation}
Similarly, $F$~is $L$-smooth if there exists a quantity~$L\geq 0$ such that, for any possible model  parameters~$\ub$ and~$\vb$,
\begin{equation}
\label{eq:mustronglyconvex}
F(\ub)  \geq F(\vb) + (\ub- \vb)^T \nabla F(\vb) + \frac{\mu}{2} \| \ub - \vb\|_2^2\,.
\end{equation}

As better detailed below,  quantities~$\mu$, $L$, and~$G$ can be computed locally at each node, by repeatedly choosing~$\ub$ and~$\vb$, and studying how the corresponding model instances perform over the local datasets. However, the bound in \Eq{bound} depends upon their global values;  e.g., the global~$G$ will be the largest of the $G$-values computed by each learning node. Additionally, it is worth noting that the local values of $\mu$, $L$, and~$G$ can be  exploited
to predict the loss improvement achieved by each individual node during local epochs, as set forth below.

\subsection{Experimental setup}

We carry out our experiments by performing an image classification task over the CIFAR-10 dataset~\cite{cifar}, containing a total of 60,000~images belonging to 10~different classes. To perform the classification, we leverage a DNN including two convolutional layers and three fully-connected ones, as per~\cite{lenet}, for a total of over 60,000 parameters. The dataset is partitioned into a {\em testing} set of 6,000~images and a {\em training} set of 54,000~ones; the latter is further partitioned into {\em local datasets} associated with the individual learning nodes.

\begin{figure}[t] \centering
\includegraphics[width=0.9\columnwidth]{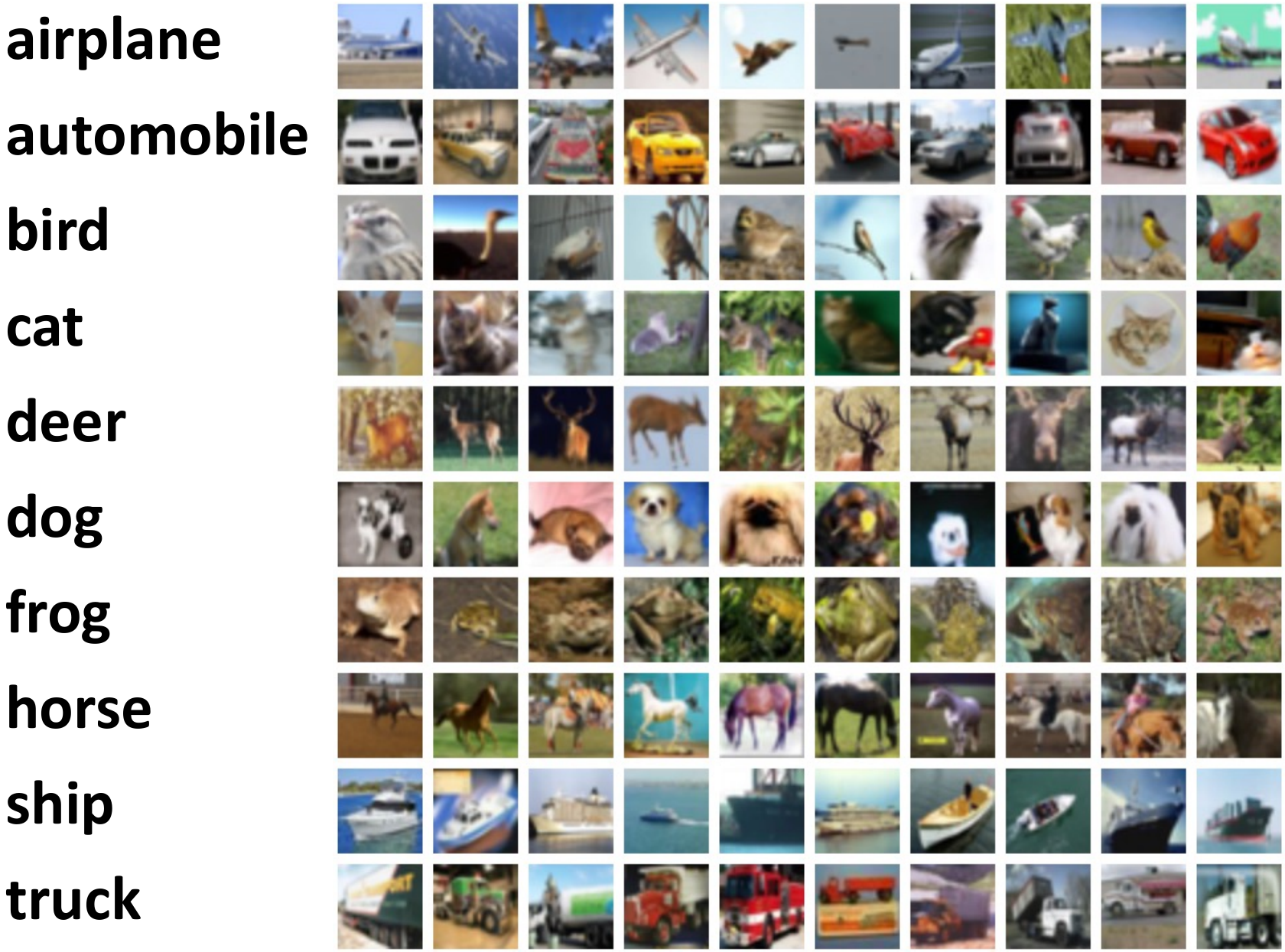} 
\caption{Classes and example images of the CIFAR-10 dataset~\cite{cifar}.
\label{fig:cifar} } %caption 
%\vspace{-4mm}
\end{figure}

At each local node, we compute~$\mu$, $L$ and~$G$ as follows:
\begin{enumerate}
    \item we extract two random sets of parameters~$\ub$ and $\vb$;
    \item we compute the resulting loss values~$F(\ub)$ and~$F(\vb)$ over the local datasets;
    \item we compute the gradients of the loss~$\nabla F(\ub)$ and~$\nabla F(\vb)$;
    \item we compute the value $m=2\frac{F(\ub)-F(\vb)+(\vb-\ub)^T\nabla F(\vb)}{\lVert\ub-\vb\rVert_2^2}$ and store it;
    \item we compute the value $g=\sqrt{\lVert F(\vb)\rVert^2}$ and store it;
    \item we repeat the above steps starting  from (1) until
a sufficiently large number of samples has been collected.
\end{enumerate}
After all samples have been collected, (i) as per \cite[Assumption~1]{li2019convergence}, $\mu$ is set to the smallest of the $m$-values; (ii) as per \cite[Assumption~2]{li2019convergence}, $L$ is set to the largest of the $m$-values, and (iii) as per \cite[Assumption~4]{li2019convergence}, $G$ is set to the largest of the $g$-values.

We compare three different learning scenarios, changing the number of nodes and the quality of their data. In the basic scenario (labelled as ``5 nodes'' in the plots shown in \Sec{results}), there are five learning nodes, each with 2,000~images drawn from the CIFAR-10 dataset. We then consider a richer scenario (labelled as ``10 nodes'' in the plots) where we double the number of learning nodes. Finally, we consider a more challenging scenario (labelled as ``5 nodes, missing class'' in the plots), where there are five learning nodes, each has 2,000~images, {\em and} all samples of one class (namely, {\em ship}) are missing from all training sets.

\section{Experimental Analysis and Main findings}
\label{sec:results}

\begin{figure*}
\centering
\subfigure[\label{fig:loss-train}]{
    \includegraphics[width=.32\textwidth]{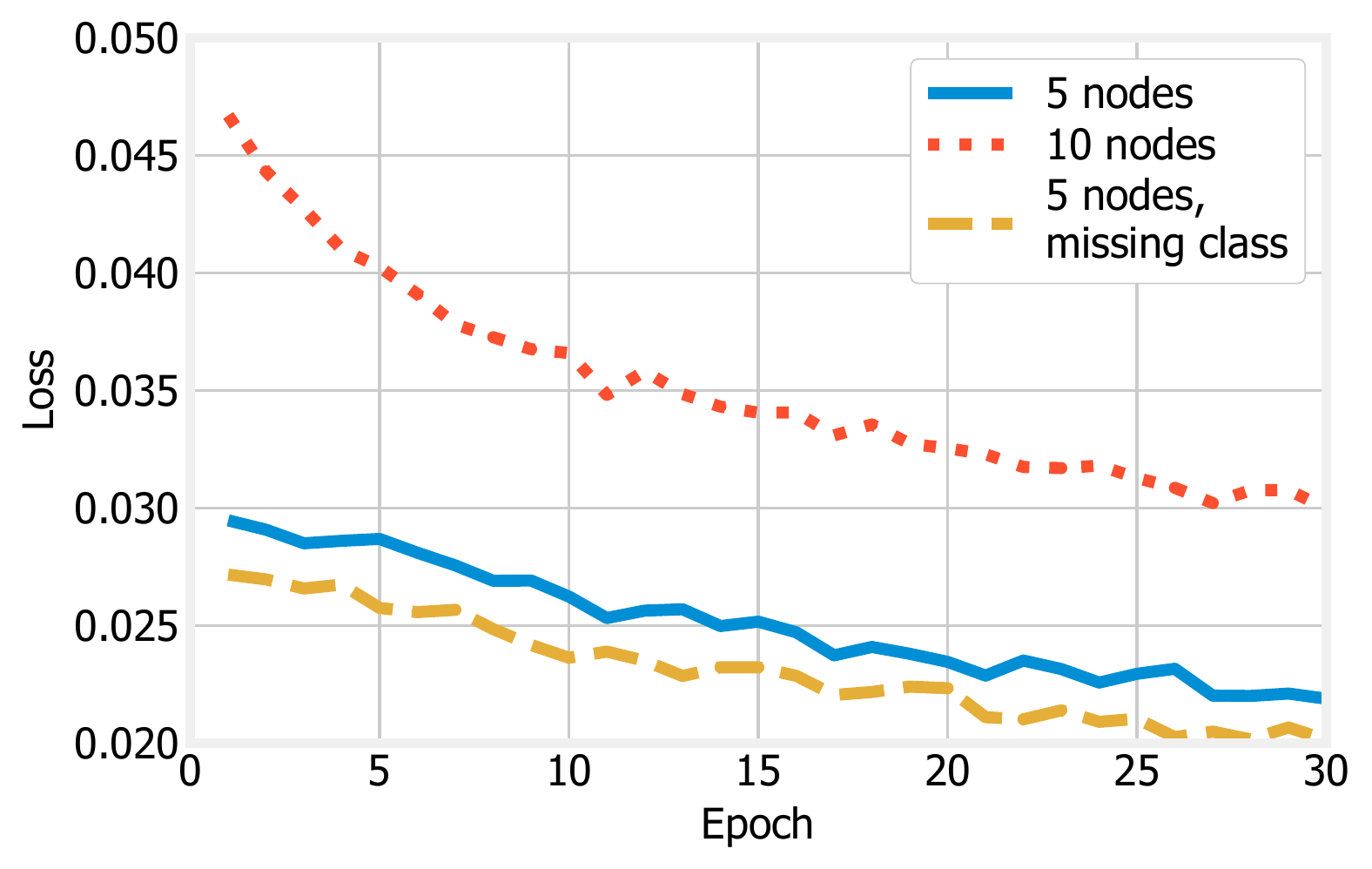}
} %subfigure
\subfigure[\label{fig:loss-test}]{
    \includegraphics[width=.32\textwidth]{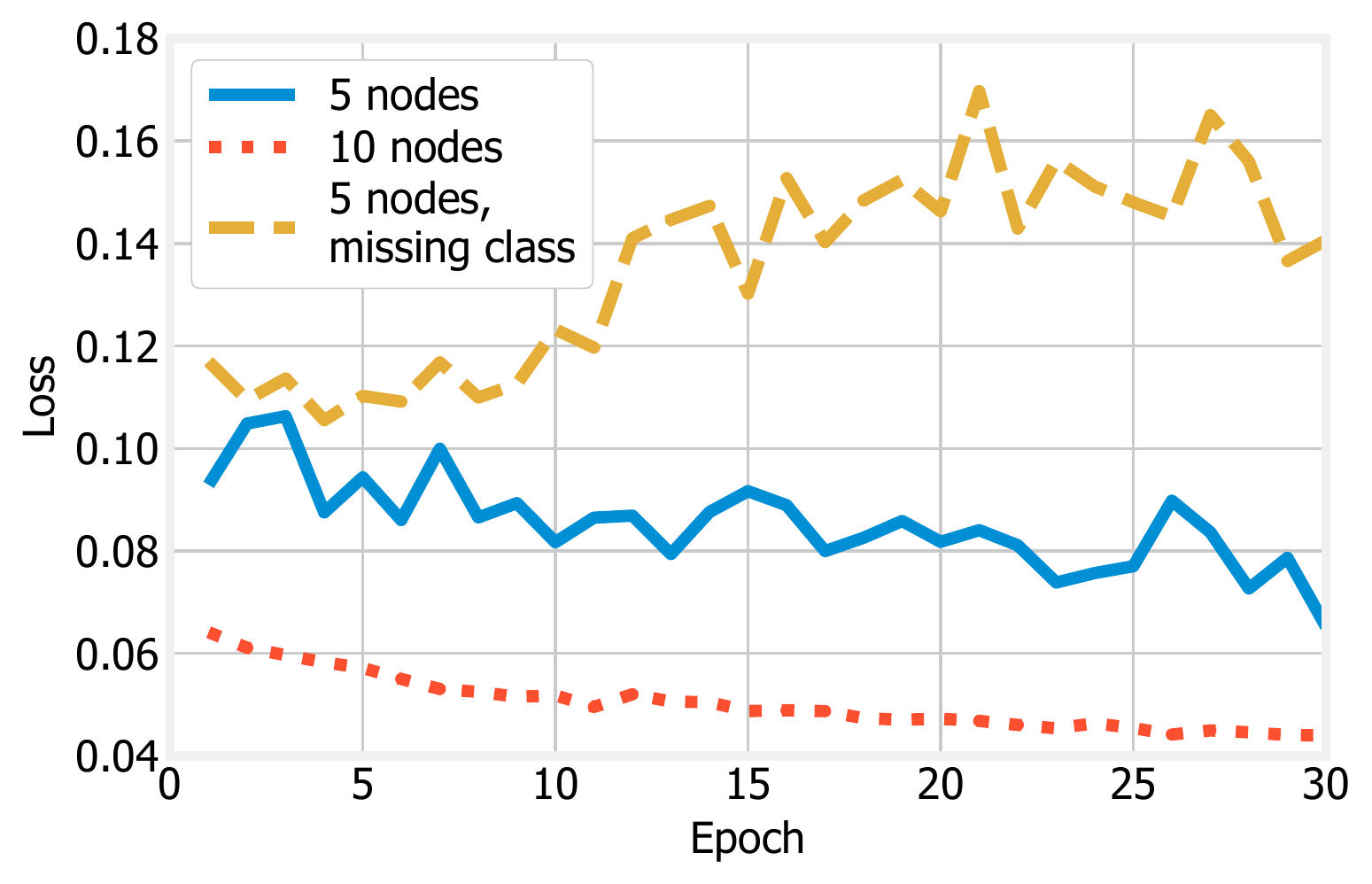}
} %subfigure
\subfigure[\label{fig:loss-bounds}]{
    \includegraphics[width=.32\textwidth]{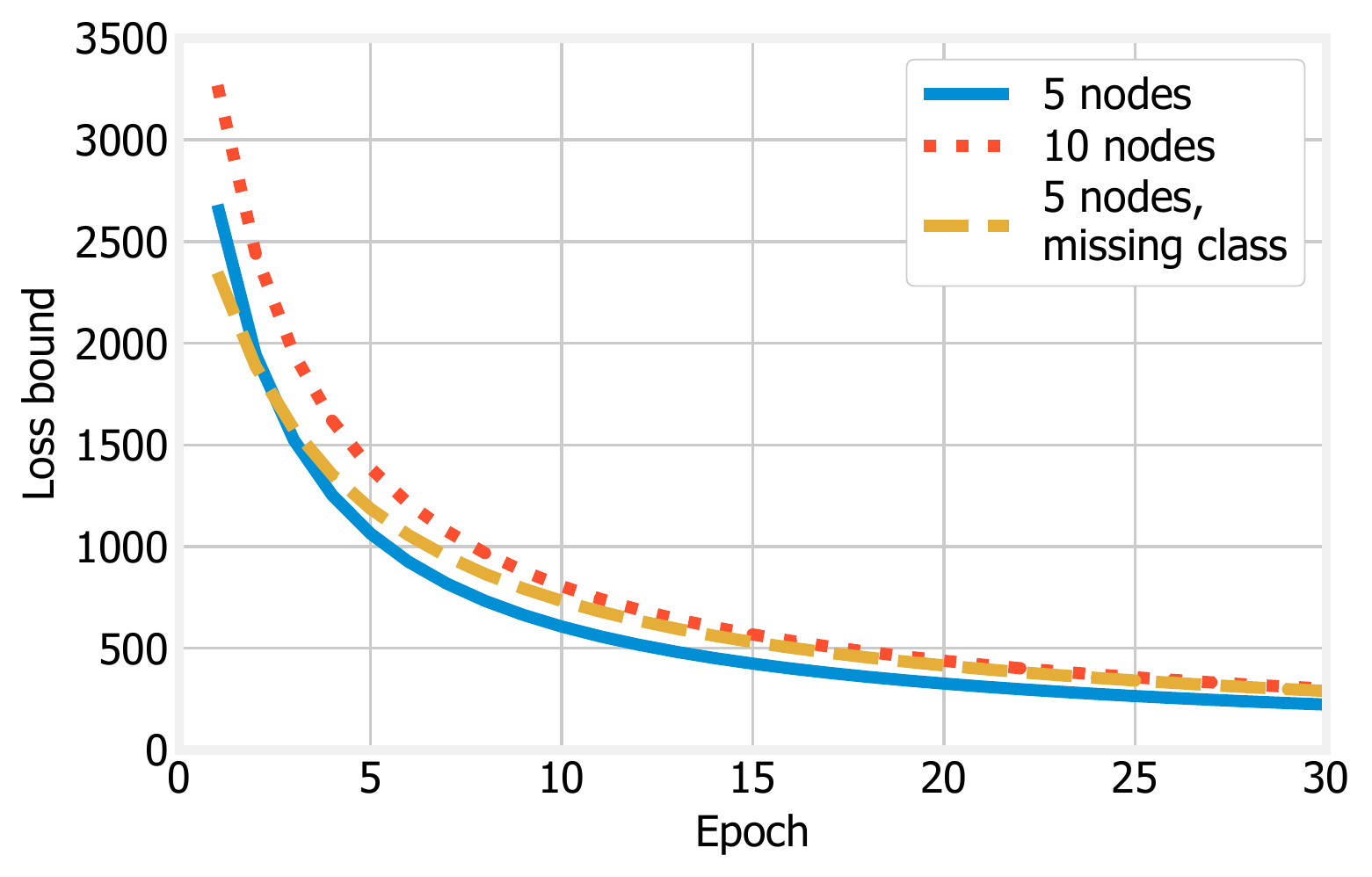}
} %subfigure
\caption{
    FL experiments: loss achieved during the training (a) and testing (b) phase; bounds thereto (c).
    \label{fig:loss}
} %caption
%\end{figure*}
%\begin{figure*}
\centering
\subfigure[\label{fig:scatter-L}]{
    \includegraphics[width=.32\textwidth]{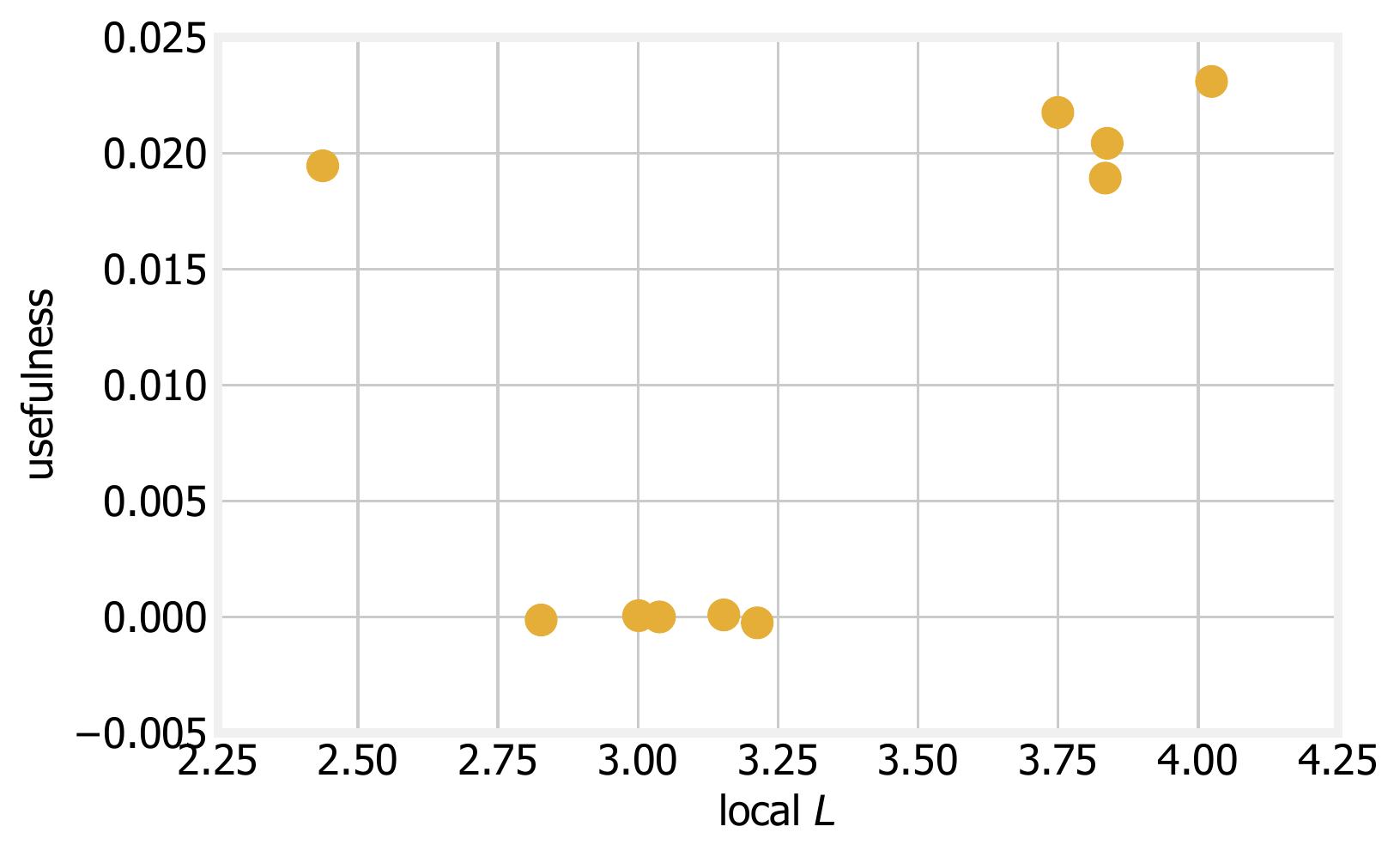}
} %subfigure
\subfigure[\label{fig:scatter-mu}]{
    \includegraphics[width=.32\textwidth]{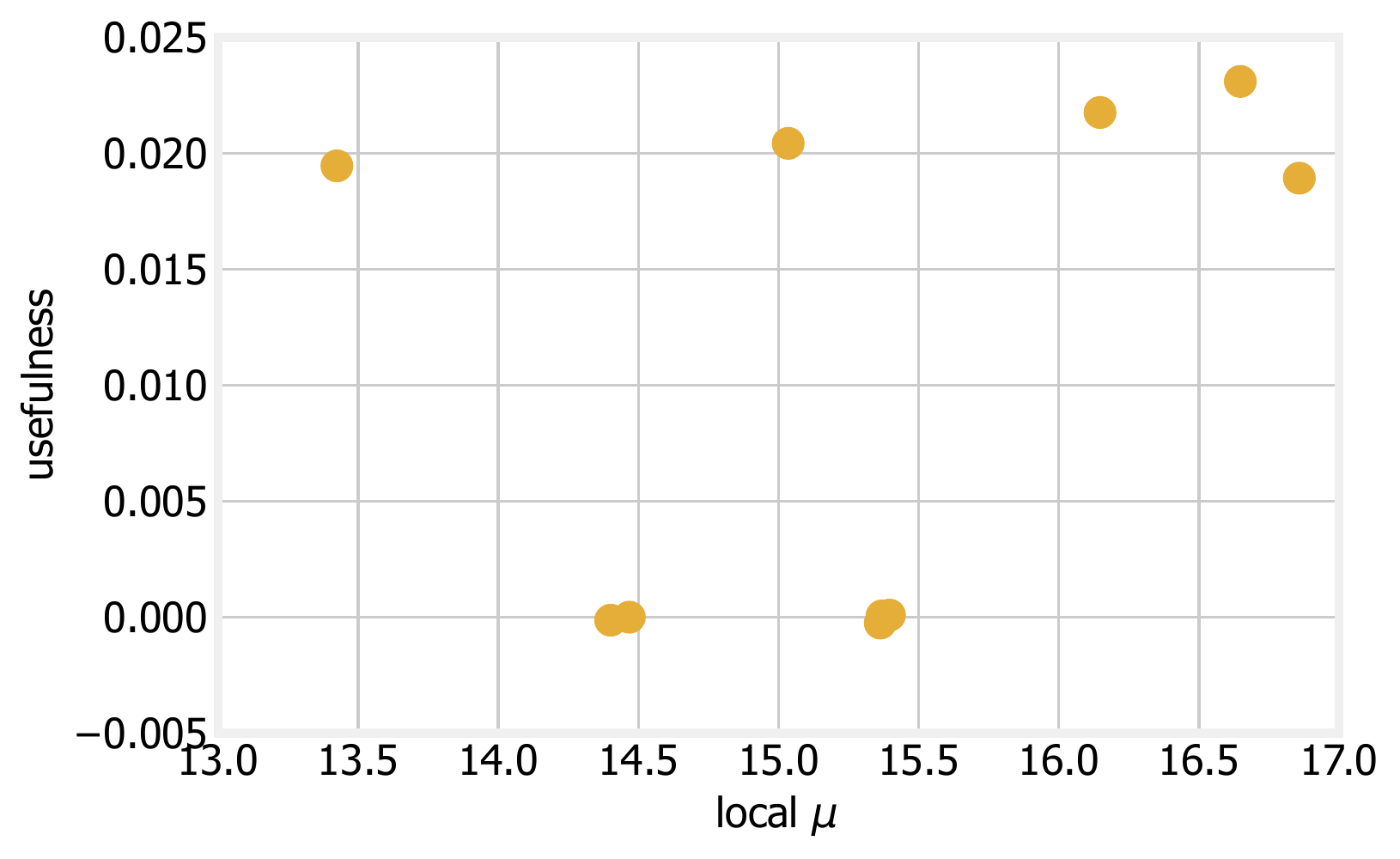}
} %subfigure
\subfigure[\label{fig:scatter-G}]{
    \includegraphics[width=.32\textwidth]{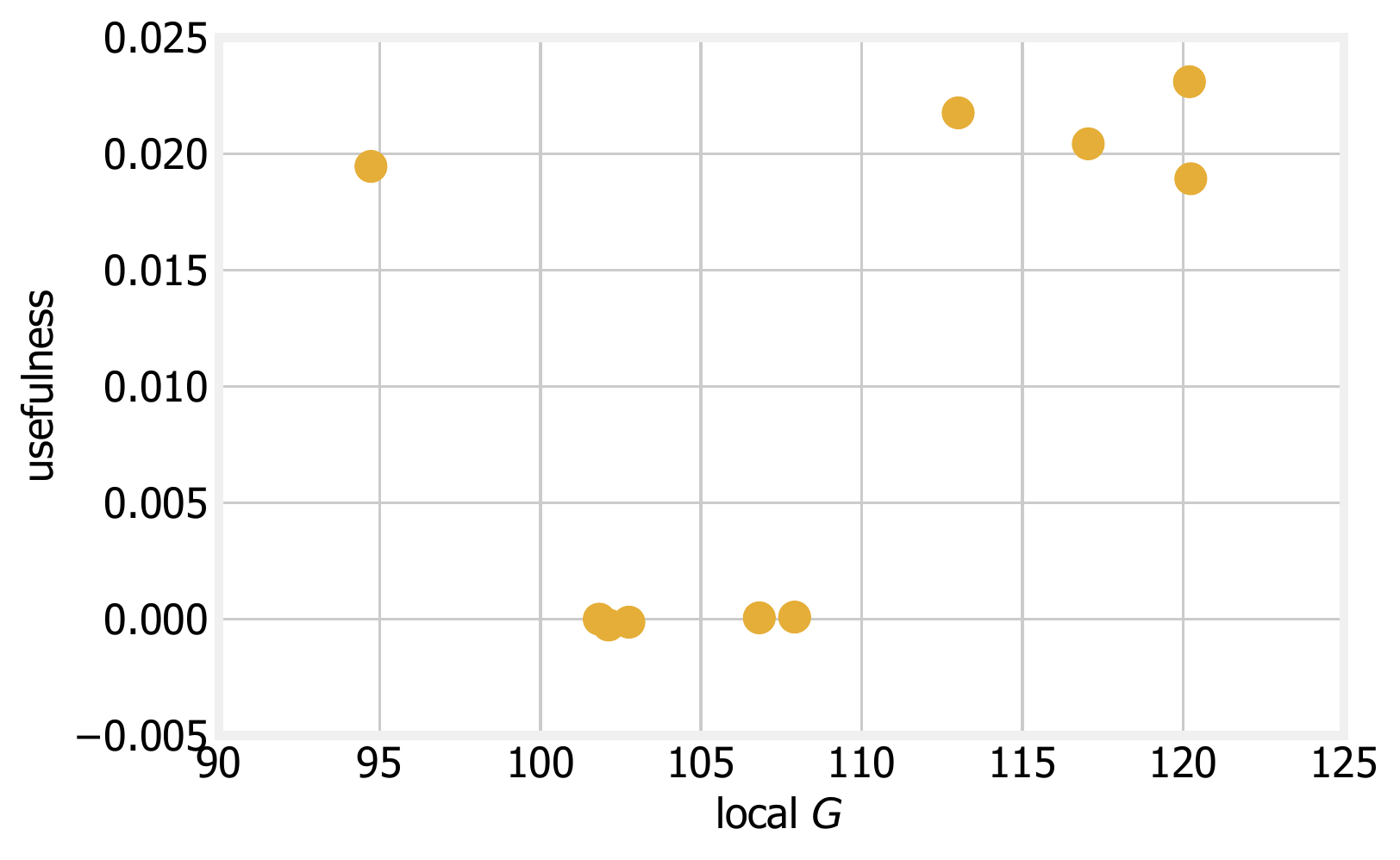}
} %subfigure
\caption{
    FL experiments: relationship between the node usefulness and the local values for the $L$ (a), $\mu$ (b) and $G$ (c) quantities.
    \label{fig:scatter}
} %caption
\vspace{-2mm}
\end{figure*}

The first aspect we are interested in is the extent to which bounds match, qualitatively and quantitatively, the behavior of the actual loss. To this end, \Fig{loss} shows the evolution of the training loss (\Fig{loss-train}) and of the testing loss (\Fig{loss-test}), and the bounds thereto (\Fig{loss-bounds}). 
Looking at \Fig{loss-train} and \Fig{loss-test} and comparing the blue solid  line and the red dotted line therein, we can observe that, as expected,  having more learning nodes increases the training loss (i.e., intuitively, it is harder to converge to a good model) but decreases the testing one (i.e., the resulting model works better with hitherto unknown data). The yellow dashed lines, describing the effect of removing a whole class from all training datasets, show very different effects on the testing and training loss.  Having fewer classes to learn makes training easier (hence, a lower training loss in \Fig{loss-train}). However, the resulting model performs very poorly over the testing set (hence, a higher loss in \Fig{loss-test}). Both these effects make intuitive sense and are routinely observed in similar scenarios.

More interestingly, \Fig{loss-bounds} depicts the loss bounds, i.e., the value of \Eq{bound}, for the three scenarios. By looking at the scale of the $y$-axis, the first thing we can notice is that bounds are orders of magnitude larger than the corresponding loss values -- which is to be expected, as bounds have to account for the worst possible conditions over {\em all} choices of~$\ub$ and~$\vb$. Perhaps more relevant, the {\em qualitative} relationship between the bounds of different scenarios follows neither the train losses in \Fig{loss-train} nor the testing losses in \Fig{loss-test}. Furthermore, the bounds provide no warning about the serious problems arising from whole categories missing in the training set (yellow lines in \Fig{loss-train} and \Fig{loss-test}).

However, a more detailed analysis surprisinly shows that, although the bounds themselves cannot be directly used to predict and improve the performance of real-world ML tasks, some of their components can be very useful. In \Fig{scatter}, we examine the relationship between the three quantities we compute to determine the bounds, i.e., $\mu$, $L$, and $G$, and the {\em usefulness} of each node within the cooperative training. The usefulness metric is defined~\cite{malandrino2021federated} as average improvement in testing loss achieved by  learning nodes during their local iterations; the underlying intuition is that nodes with a larger usefulness ``push'' the learning further during their local epochs.

It is interesting to notice how~$L$ and (to a lesser extent) $G$~have a strong correlation with node usefulness. It follows that computing {\em local} values of such quantities can significantly help  identify the nodes that are more likely to give a better contribution to the cooperative learning, a very important problem in all distributed learning scenarios. Even more importantly, computing and sharing these quantities require nodes to disclose {\em no information} about the size and quality of their dataset, which is instead required by many existing node selection schemes  and may result in privacy leakage.

Another very interesting aspect we can notice from \Fig{scatter} is that {\em higher} values of both~$L$ and~$G$ are associated with {\em higher} usefulness; however, as per \Eq{bound}, high values of both~$L$ and~$G$ make the value of the bound larger, i.e., indicate a worse learning. We can make sense of this apparent contradiction by remembering to what exactly the bound in \Eq{bound} refers, that is, the {\em training} loss. Intuitively, a good way to obtain a low training loss is to have a small training dataset, with samples that are not too different from each other. Indeed, moving to a degenerate scenario, a dataset with only {\em one} class represented therein can be learned with zero training loss by a DNN always predicting that class. On the other hand, generalization (hence, good performance over the testing set) requires larger datasets of higher quality, which may require more training epochs, thus, incur a higher training loss.

This discrepancy also points at a higher-level aspect that it is essential to keep in mind, in order to understand and leverage convergence results:  bounds are based upon the analysis of the behavior of the stochastic gradient descent (SGD) {\em optimization} algorithm. While optimization is a fundamental part of ML, ML is much more than optimization; therefore, there are many aspects of ML that convergence bounds, by their nature, cannot capture.

\begin{figure}[t] \centering
\includegraphics[width=.8\columnwidth]{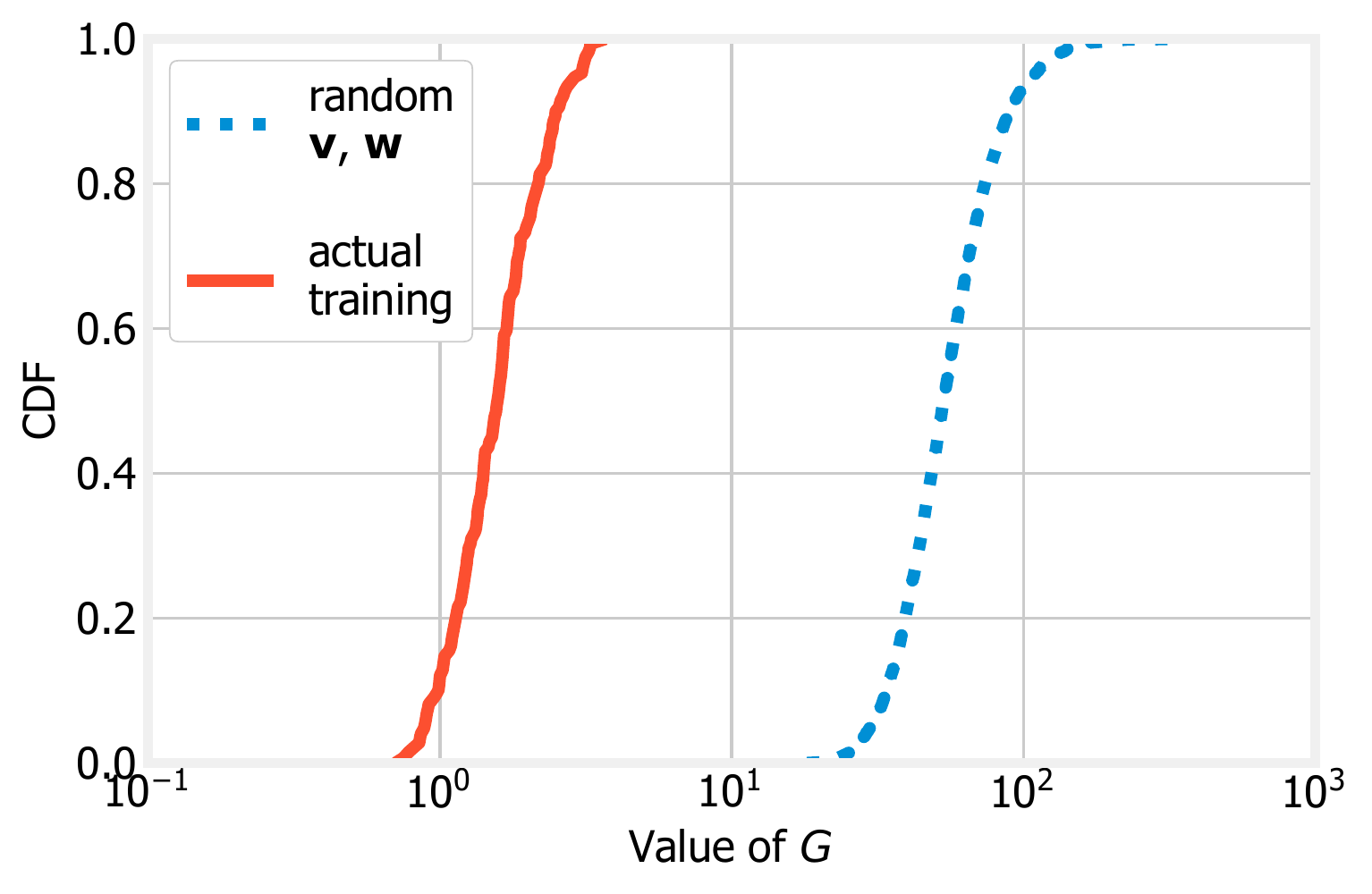} 
\caption{FL experiments: distribution of the values of $G$ measured through random~$\ub$ and~$\vb$ values (blue) and during actual training (red).
\label{fig:g-cdf} } %caption 
\vspace{-4mm}
\end{figure}

A further example is shown in \Fig{g-cdf}, presenting the cumulative density function (CDF) of the quantity~$G$ obtained by selecting random~$\ub$ and~$\vb$ values (blue dotted curve in the plot) and the $G$-values observed during actual training (red solid curve). We can immediately see that the values obtained from random~$\ub$ and~$\vb$ values are over an order of magnitude larger than those actually observed during training. Recalling that bounds must hold for even the largest possible values of $G$, i.e., the top point of the blue curve, this explains why the bounds in \Fig{loss-bounds} are as loose as they are.

This also ties with our earlier remark about the intrinsic limit of convergence studies, i.e., there are aspects of ML that simply cannot be captured by  convergence studies. In the case of \Fig{g-cdf}, the gradients (hence, the $G$-values) encountered during training are relatively small precisely because a lot of effort and research in the field of ML, e.g., DNN initialization schemes, learning rate adaptation algorithms, etc., have been devoted to keeping gradients low. In other words, one may say that convergence bounds capture the optimization aspect of ML, but not the many techniques used in ML to make the optimization perform better.

\section{Conclusion and future work}
\label{sec:conclusion}

In the context of distributed learning, it is of paramount importance to estimate  how many training epochs will be needed to reach the target learning performance; this, in turn, depends upon how much the loss function can be reduced in a single epoch. Convergence analysis results, based on the performance and behavior of SGD, are a very valuable tool to estimate this important quantity {\em a priori}, that is, before actually starting to train the network.

In this work, we have leveraged a set of experiments based on federated learning to (i) assess to which extent the bounds reflect, qualitatively and quantitatively, the behavior of actual DNN training, and (ii) whether the quantities appearing in the bounds can be leveraged to improve the performance of distributed learning. 
Our major findings can be summarized as follows:
\begin{itemize}
    \item the full convergence bounds have only a loose relationship with the qualitative and quantitative evolution of the testing and training losses;
    \item nonetheless, the {\em quantities appearing therein can be very useful} to identify the learning nodes where local updates yield the largest loss reduction.
\end{itemize}
The latter metric is linked to how effectively each node can contribute to the learning process~\cite{malandrino2021federated}, hence, it is also useful towards more effective node selection.

Our results also highlight a fundamental feature of all convergence studies, i.e., that they can well capture the behavior of the optimization component of ML, while it is much harder for them to account for the techniques used in ML to improve the performance of optimization. In spite of this inherent limitation, as noted above, theoretical convergence studies can be very valuable in identifying the most suitable nodes to participate in the distributed ML process, thereby improving the performance of the learning itself.

Future work will focus on leveraging such insights to build a concrete algorithm for the selection of learning nodes, and evaluate its performance over a wide set of datasets and DNN architectures.

\bibliographystyle{IEEEtran}
\bibliography{refs}

\end{document}